**Enhanced Elevated-Temperature Strength in Refractory Complex Concentrated Alloys via Temperature-Induced Transition from Screw-to-Edge Dislocation Control**

*Tamanna Zakia[1], Ayeman Nahin[1], Dunji Yu[2], Jacob Pustelnik[1], Juntan Li[3], Mason Kincheloe[1], Lia Amalia[2,3], Yan Chen[2], Peter K. Liaw[3], Haixuan Xu[3], Mingwei Zhang[1]*

[1]Department of Materials Science and Engineering, University of California, One Shields Ave., Davis, CA, 95616, USA.

[2] Neutron Scattering Division, Oak Ridge National Laboratory, Oak Ridge, TN, 37830

[3] Department of Materials Science and Engineering, University of Tennessee, Knoxville, TN, 37996, USA

Corresponding Author:

Mingwei Zhang, Assistant Professor

Department of Materials Science and Engineering

University of California, Davis

Email: mwwzhang@ucdavis.edu

**Abstract**

Refractory complex concentrated alloys (RCCAs) show promise for high-temperature applications but often lose strength due to screw-dislocation-controlled plasticity. We demonstrate a temperature-driven transition from screw- to edge-dislocation-controlled deformation in a single-phase NbTaTiV RCCA. Tensile tests from 298–1573 K reveal a pronounced intermediate-temperature strength plateau and yield strengths surpassing other ductile RCCAs and the Ni-based superalloy CMSX-4 above 1273 K. In-situ neutron diffraction, TEM, and molecular dynamics identify a crossover near ~900 K, where edge dislocation glide stabilized by V-induced lattice distortion dominates, enabling enhanced strength retention and a clear design strategy for ultrahigh-temperature applications.

**Keywords:** Refractory complex concentrated alloys, deformation mechanisms, dislocation, high temperature strength.

**Impact Statement:** This work uncovers a temperature-driven transition from screw- to edge-dislocation-controlled strength in NbTaTiV, enabling exceptional high-temperature strength retention and defining a new lattice-distortion-based strategy for ultrahigh-temperature alloy design.

## 1. Introduction

Our growing demands in space and energy systems require structural materials that operate beyond the temperature limits of Ni-based superalloys [1]. Refractory complex concentrated alloys (RCCAs) have recently emerged as promising candidates, defined by compositions containing high concentrations of at least three refractory elements [2]-[3]. RCCAs have inherently high melting points ($T_m$) due to the incorporation of Group IV-VI refractory elements, enabling higher operating temperatures than Ni-based superalloys (~1600 K) and, in some cases, excellent high-temperature strength retention. For example, MoNbTaVW retains >400 MPa compressive strength at 1873 K [4], but like many Group VI–rich RCCAs (Cr, Mo, W), it shows poor tensile ductility due to intrinsic and grain boundary brittleness, limiting its practical use [5]-[7]. To improve tensile ductility, Group VI-free RCCAs such as the HfNbTaTiZr (Senkov) alloy and its derivatives were developed and widely studied [8]-[16]. While they show good ductility, fracture toughness, and room-temperature strength, their strength drops rapidly below 1273 K, making them less competitive with Ni-based superalloys. This degradation arises from thermally activated screw-dislocation-mediated mechanisms, such as overcoming Peierls barriers, cross-kink unpinning, and jog formation, which were supported by TEM observations. Therefore, achieving a transition to less temperature-sensitive deformation mechanisms is critical for better strength retention in RCCAs.

Recent work by Lee et al. [23] identified NbTaTiV as a promising candidate to achieve such transition, showing a high compressive strength up to 1173 K. Vanadium introduces a large atomic size misfit (~8% smaller radius) [24] that can preferentially impede edge dislocation motion and is proposed to shift high-temperature strength control from screw- to edge-dislocation-controlled mechanisms with weaker temperature sensitivity [17], [23], [25]. However, direct experimental validation of this mechanistic shift is still limited. Here, we provide comprehensive experimental and computational evidence for a temperature-driven transition from screw- to edge-dislocation-controlled glide in NbTaTiV using high-temperature tensile testing, in-situ neutron scattering, TEM, and molecular dynamics simulations.

## 2. Materials and Methods

The equiatomic NbTaTiV alloy was fabricated by vacuum arc melting from ultrahigh-purity elements. The chamber was evacuated to $<5 \times 10^{-5}$ torr and backfilled with ultrahigh-purity argon, with titanium getters used to minimize oxygen contamination. To ensure compositional homogeneity, the alloy button was flipped and remelted at least five times. As-cast buttons were cold rolled to a thickness of 1 mm using an IRM2050 2-hi rolling mill, and dog-bone tensile specimens were machined by electrical discharge machining. Specimens were wrapped in tantalum foil and annealed at 1473 K for 3 h under high vacuum, followed by air cooling at ~100 K/min using a sliding-tube furnace to achieve full homogenization and recrystallization. Oxygen and nitrogen contents were measured by inert gas fusion (LECO OHN836) at IMR Test Labs (Portland, OR).

Room-temperature tensile tests were conducted on an MTS Model 810 servohydraulic machine, while elevated-temperature tests were performed on an Instron 1331 hydraulic testing machine equipped with a Mo-element vacuum furnace (up to 1873 K). High-temperature tests were carried out from 573 to 1573 K at 200 K intervals under high vacuum ($<5 \times 10^{-5}$ torr). Temperature was monitored using a W–Re thermocouple near the gauge section (±2 K). Samples were heated at 15

K/min, held for 5 min, and deformed at a strain rate of $5 \times 10^{-4}$ $s^{-1}$. Tests were interrupted below 2% strain for TEM analysis.

TEM characterization was performed at 200 kV using a JEOL 2100F (UC Davis) and an FEI Tecnai F20 UT (National Center for Electron Microscopy, Lawrence Berkeley National Laboratory). SEM and EDS analyses were conducted using a ThermoFisher Quattro S Environmental SEM (UC Davis). Samples were mechanically ground to 1200 grit and electropolished in 10 vol% $H_2SO_4$ in methanol at −20 °C.

In-situ neutron diffraction experiments were performed on the VULCAN Engineering Diffractometer at the Spallation Neutron Source (Oak Ridge National Laboratory). Uniaxial tensile tests with in-situ heating were conducted on an MTS load frame at a strain rate of $5 \times 10^{-6}$ $s^{-1}$ at 298, 673, 873, and 1173 K. Dog-bone specimens (100 mm total length, 15 mm gauge length, 2.5–3 mm thickness) were used, with a K-type thermocouple spot-welded to the gauge section. Tests were interrupted at 2% strain and air cooled to room temperature. Simultaneous neutron data collection enabled extraction of elastic constants and dislocation analysis from diffraction peak broadening.

MD simulations were performed using the Large-scale Atomic/Molecular Massively Parallel Simulator (LAMMPS) [26] with a modified Embedded Atom Method (MEAM) potential for NbTaTiV [27]. Edge and screw $a/2\langle 111\rangle$ dislocations on $\{110\}$ planes were with randomly distributed atomic species. Simulation cell dimensions were $x[11\bar{2}]$, $y[111]$, and $z[1\bar{1}0]$ for an edge dislocation and $x[111]$, $y[11\bar{2}]$, and $z[1\bar{1}0]$ for a screw dislocation. The simulation box contained ~2×10$^5$ atoms (≈80×250×160 Å) using a 1 fs timestep and a Nosé–Hoover thermostat [28] over 300–1200 K.

### 3. Results and Discussion

SEM backscattered electron micrograph (SEM-BSE) in **Figure 1A** reveals full homogenization and recrystallization of the as-rolled NbTaTiV tensile specimen. The grain boundaries are clean, with no evidence of phase decomposition. **Figure 1B** presents the corresponding EDS elemental maps and confirms a uniform spatial distribution of Nb, Ta, Ti and V with no sign of elemental segregation. The average grain size was determined by the linear intercept method to be 62 ± 9 μm. Neutron diffraction patterns obtained from the undeformed sample are shown in **Figure 1C.** The diffraction intensities confirm the formation of a single, stable BCC phase in NbTaTiV sample. Minor FCC peaks appeared in the diffraction pattern are from the K-type thermocouple used for in-situ heating/tension. The lattice parameter determined from the neutron diffraction measurement was 3.24 Å. The grain size, lattice parameter, and the actual chemical compositions of the alloy obtained from EDS and inert gas fusion are summarized in **Table 1**.

**Table 1 List of nominal and actual compositions, grain sizes, neutron measured lattice constants, and inert gas fusion O and N interstitial contents (in atomic ppm).**

| Nominal Composition | Grain Size (μm) | Experimental Lattice Constants (Å) | EDS Composition (Atomic %) | | | | Inert Gas Fusion Composition (Atomic ppm) | |
|---|---|---|---|---|---|---|---|---|
| | | | Nb | Ta | Ti | V | O | N |
| NbTaTiV | 62 ± 9 | 3.2393 | 23.74 | 22.79 | 27.86 | 25.61 | 5333.4 | 531.76 |

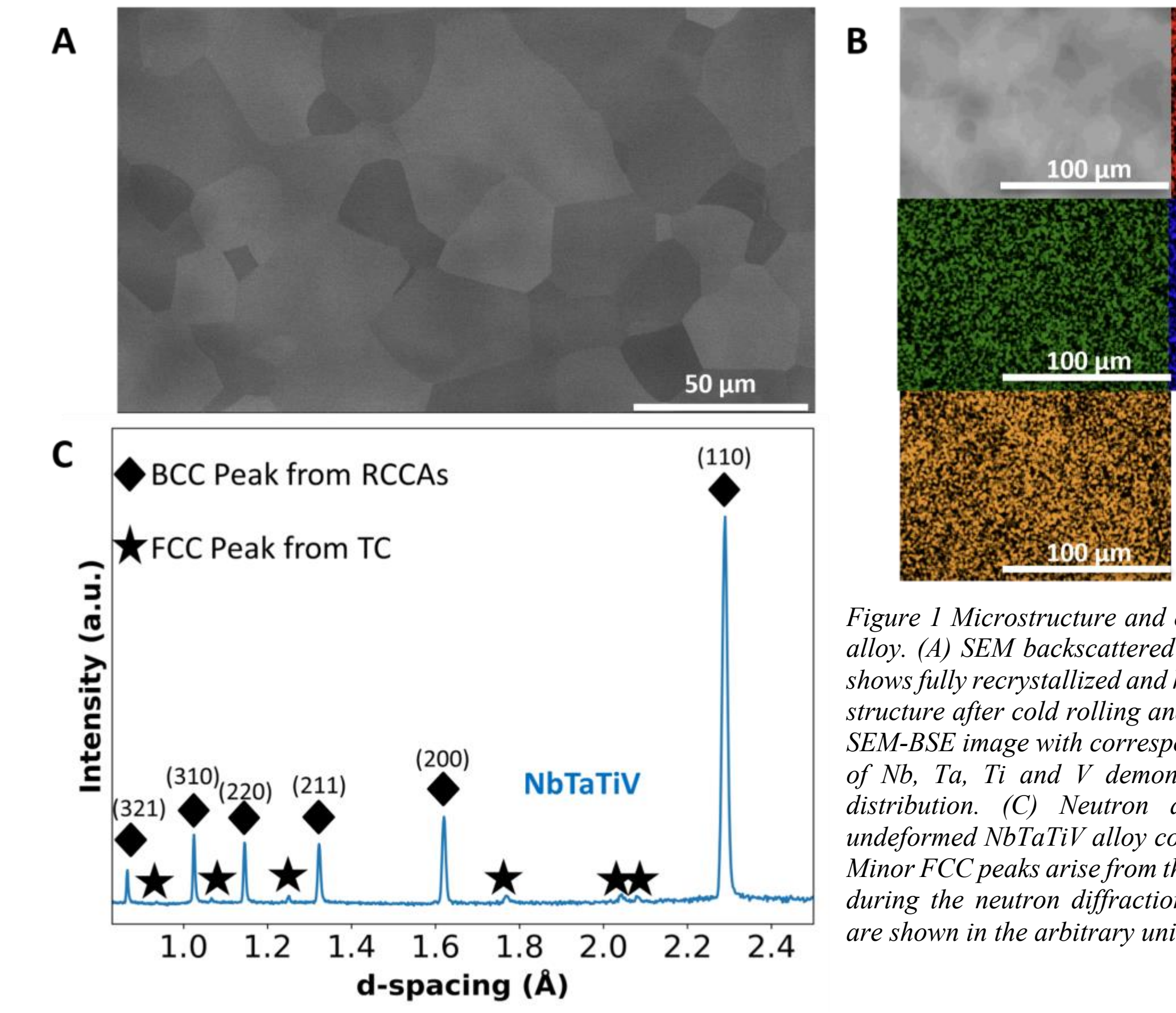

*Figure 1 Microstructure and elemental maps of NbTaTiV alloy. (A) SEM backscattered electron (SEM-BSE) image shows fully recrystallized and homogenized equiaxed grain structure after cold rolling and subsequent annealing. (B) SEM-BSE image with corresponding EDS elemental maps of Nb, Ta, Ti and V demonstrating uniform elemental distribution. (C) Neutron diffraction pattern of the undeformed NbTaTiV alloy confirms a stable BCC phase. Minor FCC peaks arise from the K-type thermocouple used during the neutron diffraction measurements. Intensities are shown in the arbitrary units.*

In **Figure 2**, the 0.2% offset tensile yield strength of the NbTaTiV alloy as a function of temperature reveals three distinct deformation regimes: a low temperature thermal sensitive regime, an intermediate temperature plateau regime, and a high-temperature thermally activated regime. Up to 573 K (~ 0.2 $T_m$), the yield strength exhibits a sharp decrease with increasing temperature. In this region, plastic deformation has been reported to be dominated by short range barriers such as the Peierls stress and kink pair nucleation of screw dislocation [29]. A relatively temperature-independent plateau region at intermediate temperatures (~0.2-0.4 $T_m$) suggests a possible transition in the dominant strengthening mechanisms. Beyond 1173 K (~0.45) $T_m$, yield strength decreases rapidly again, indicating the onset of diffusion-mediated mechanisms, such as dislocation climb-controlled plasticity and/or dynamic recrystallization [30], [31].

**Significantly better strength retention at elevated temperatures for NbTaTiV** was observed in comparison with other ductile RCCAs (**Figure 2(A)**), such as NbTaTi [16], $Nb_{45}Ta_{25}Ti_{15}Hf_{15}$ [15], and HfNbTaTiZr [8]-[9], whose strengths are previously shown to be controlled by the glide of screw dislocations. At 1173 K, the yield strength of NbTaTiV is approximately twice that of $Nb_{45}Ta_{25}Ti_{15}Hf_{15}$ and HfNbTaTiZr, and nearly three times that of NbTaTi. More importantly, NbTaTiV can show superior yield strength compared to a leading single-crystal Ni-based superalloy, CMSX-4, beyond a temperature of 1273 K. Although the widely used CMSX-4 displays peak yield strength at ~1000 K, which is consistent with the strengthening behavior of Ni-based superalloys [32], the yield strength of CMSX-4 decreases sharply beyond this temperature due to the loss of $\gamma'$ strengthening. In contrast, NbTaTiV maintains excellent strength retention beyond 1273 K, demonstrating its enhanced high-temperature mechanical stability. Its superior tensile yield strength up to 1573 K, beyond the maximum operating temperature of Ni-based superalloys, represents a milestone in the development of ductile, single-phase RCCAs. This performance highlights the advantage of concentrated solid-solution strengthening driven by V-induced lattice distortion, which can offer more favorable ultrahigh-temperature properties than precipitation strengthening that becomes vulnerable when the temperature approaches the solvus temperature of the intermetallic phase.

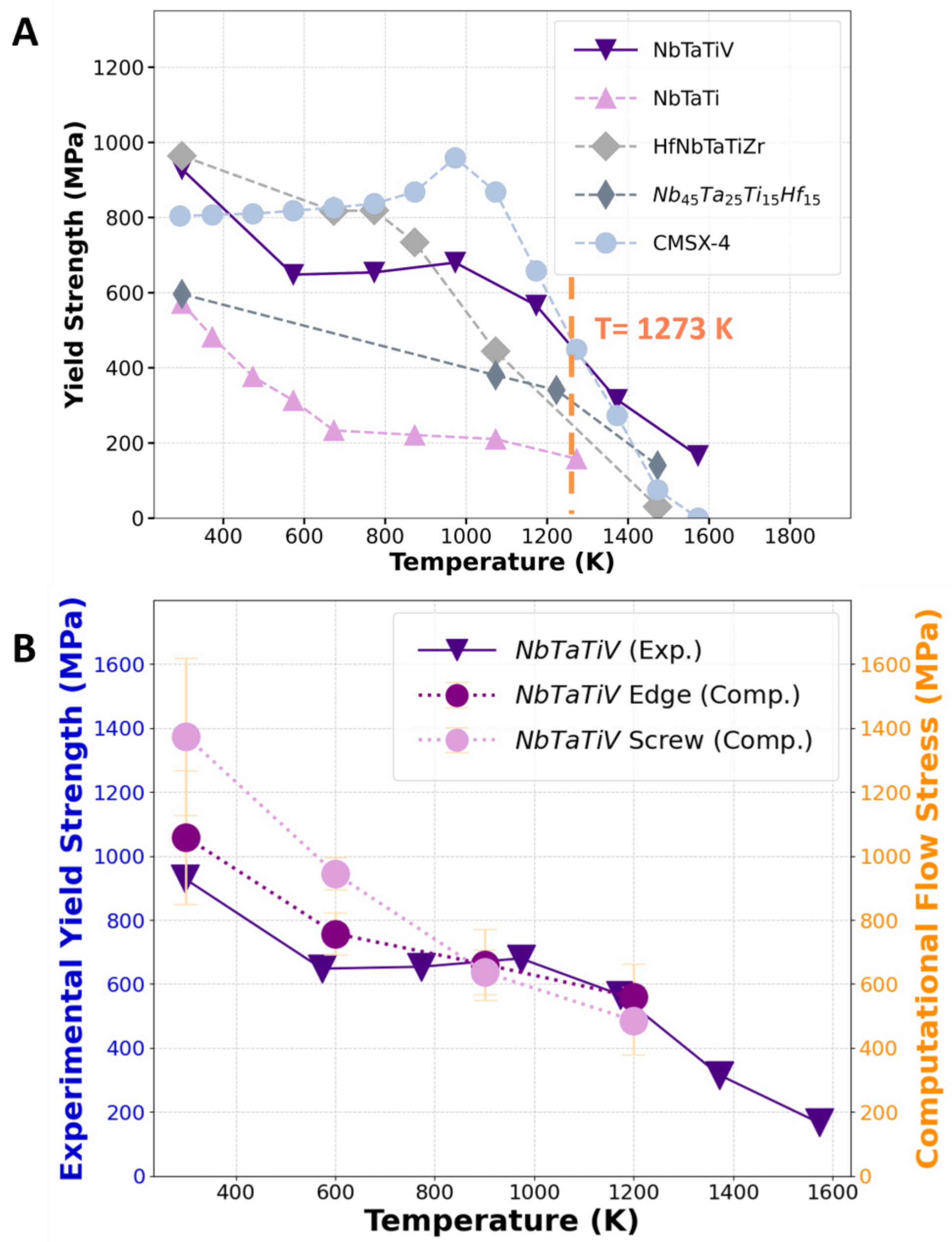

***Figure 2.*** *Temperature dependence of the 0.2% offset tensile yield strength of the NbTaTiV alloy. (A) Comparison of NbTaTiV with NbTaTi, HfNbTaTiZr,, $Nb_{45}Ta_{25}Ti_{15}Hf_{15}$, and the Ni-based superalloy CMSX-4, highlighting the superior strength retention of NbTaTiV beyond 1273 K. (B) Experimental yield strength of NbTaTiV overlaid with molecular dynamics-simulated dislocation flow stresses for edge and screw dislocation glide, shows a crossover from screw to edge control near ~900 K.*

It is evident from **Fig. 2A** that intermediate-temperature plateaus are widespread in other RCCAs, especially for NbTaTi, which have been attributed to athermal average atomic size misfits and elastic modulus mismatches by Coury et al. [16]. However, in NbTaTi, the insufficient atomic size misfit leads to persistent screw-dislocation-dominated deformation at elevated temperatures, limiting its competitiveness in high-temperature strength. On the other hand, MD simulations of NbTaTiV (**Fig. 2B**) reveal a temperature-dependent crossover in glide barriers from screw- to edge-dislocation control at ~900 K, coinciding with the experimentally observed yield-strength plateau. The plateau regime can be partially attributed to this mechanistic transition with increasing temperatures, where edge dislocations exhibit a weaker temperature dependence of flow stress compared to screw dislocations. However, this screw to edge dislocation transition cannot fully rationalize the athermal plateau, suggesting additional strengthening mechanisms such as chemical short-range ordering (including ordered oxygen complexes) [33]-[34], cross-core diffusion [35], [36], as well as superjog nucleation [37] recently proposed for RCCAs.

To experimentally identify the dominant dislocation character that governs the plastic deformation in the near-yield region, tensile tests at a number of temperatures were interrupted at strains below 2%, followed by transmission electron microscopy (TEM) analyses. TEM observations of the specimen deformed at 298 K and 573 K (**Figure 3A-B**) reveal long, straight screw dislocations aligned along {110} slip planes. The g-dot-b analysis that determines the screw character for these dislocations at room temperature is given **Fig. S1**.

At the intermediate temperature of 773 K (**Figure 3 C**), TEM shows straight screw dislocations along {110} planes accompanied by substantial debris formation (vacancy or interstitial loops). These observations indicate that cross-slip and cross-kink unpinning events become increasingly dominant in this temperature range, which are widely reported in screw-dominated RCCAs at elevated temperatures [22], [38]. This contrasts with 298 K and 573 K deformation, where debris is largely missing, indicating that strength is likely controlled by screw dislocations overcoming the Peierls barrier. Moreover, along with straight screw dislocations, curvy mixed dislocations begin to appear as well suggesting the gradual transition in the dominant dislocation character. At higher temperatures, TEM observations at 973 K (**Figure 3D**) and STEM observations at 1173 K (**Figure 3E**) reveal a uniform distribution of curved and tangled dislocations across the overall microstructure. This indicates that plastic deformation is increasingly governed by the edge components at temperatures beyond 773 K. The g-dot-b analysis that determines the non-screw character for dislocations at 1173 K is given **Fig. S2**. Based on these evidence, 773 K represents a temperature where both screw and edge dislocations contribute to plastic deformation, whereas by 973 K the transition to edge-dominated deformation is essentially complete. Consequently, the transition temperature can be inferred to lie between 773 K and 973 K, which is in good agreement with the MD-predicted crossover temperature of ~900 K.

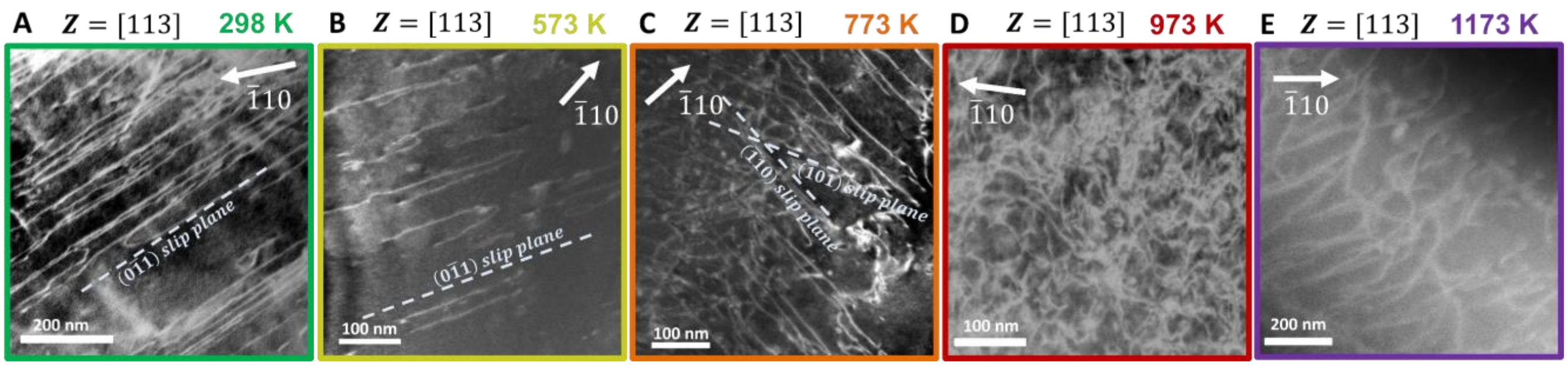

*Figure 3 TEM images of specimens deformed below 2% strain at (A) 298 K, (B) 573 K, (C) 773 K, (D) 973 K, and (E) 1173 K. (A) Long, straight screw dislocations aligned on {110} slip planes dominate at 273 K. (B)-(C) At 573 K and 773 K, straight screw dislocations on {110} planes are evident; at 773 K, curved mixed dislocations begin to appear. (D)-(E) uniformly distributed, curved, and tangled dislocations, indicative of edge- or mixed-charactered dislocations start to dominate plasticity at 973 K and 1173 K.*

For a more in-depth analysis of the dislocation character at different temperatures, in-situ neutron diffraction experiment has been performed on the NbTaTiV alloys at 298 K, 673 K, 873 K, and 1173 K following the method developed by Lee at al. [39]. **Figure 4(A)-(D)** presents the modified Williamson-Hall plot which considers the grain size and the strain effects.

$$\Delta K^2 = \left(\frac{0.9}{D}\right)^2 + \frac{\pi A^2 b^2}{2} \rho \left(K^2 C_{hkl}\right) + O\left(K^2 C_{hkl}\right)^2 \qquad (1)$$

Here, $\Delta K$ is the diffraction peak broadening, $D$ is the grain size, $A$ is the effective outer cutoff radius of dislocations, $b$ is the magnitude of the Burgers vector, $\rho$ is the dislocation density, $K = \frac{1}{d}$ is the inverse d-spacing, $O$ is a noninterpreted higher order term, and $C_{hkl}$ represents the dislocation contrast factor. $C_{hkl}$ can be expressed as

$$C_{hkl} = C_{h00}\left[1 - q\left(\frac{h^2k^2 + k^2l^2 + h^2l^2}{(h^2 + k^2 + l^2)^2}\right)\right] \qquad (2)$$

$C_{h00}$ is the contrast factor for the $[h00]$ reflection, and $q$ is a parameter which links $C_{h00}$ with $C_{hkl}$ [40]. The values of $q$ and $C_{h00}$ for NbTaTiV have previously been determined as 2.2244 and 0.2415 for a pure <111> screw dislocation, and 0.0673 and 0.1937 for a pure {110}<111> edge dislocation [39]. Values of $C_{hkl}$ can be determined by the ANIZC software for both screw and edge dislocation corresponding to their respective glide planes and glide direction [41]. Neutron diffraction peak broadening $\Delta K$ is calculated from

$$\Delta K = -\frac{\Delta d}{d^2} \qquad (3)$$

Here, $d$ (nm) is the interplanar spacing of the diffraction planes and $\Delta d$ is the full width at half maximum (FWHM) obtained from single-peak fitting of neutron diffraction data after subtracting the instrument peak broadening.

Modified Williamson-Hall plots of the NbTaTiV alloy at different temperatures are presented in **Figure 4(A)-(D),** where $\Delta K^2$ is linear fitted as a function of $K^2 C_{hkl}$ for different grain orientations using $C_{hkl}$ values corresponding to $\vec{b} = \frac{a}{2} < 111 >$ type screw and edge dislocations. At 298 K, the modified Williamson-Hall analysis shows an excellent linear fit for screw dislocations, whereas the edge dislocation fit is comparatively poor. In contrast, at 1173 K, the screw dislocation fit deteriorates substantially and the edge dislocation fit improves remarkably and exhibits excellent linearity. At intermediate temperatures (673 K and 873 K), the relative improvement of the edge fitting and the concurrent degradation of the screw fitting indicate a gradual transition rather than an abrupt shift. **Figure 4(E)** summarizes this evolution by plotting the goodness of the linear fits, $R^2$ for screw and edge dislocation as a function of temperature. The goodness of the edge dislocation fitting increases with temperature, whereas the goodness of fitting for screw dislocations decreases correspondingly. The opposing trend of screw and edge dislocation fit suggests that the plastic deformation in the near-yield region is governed by a transition of the dominant dislocation character at elevated temperatures.

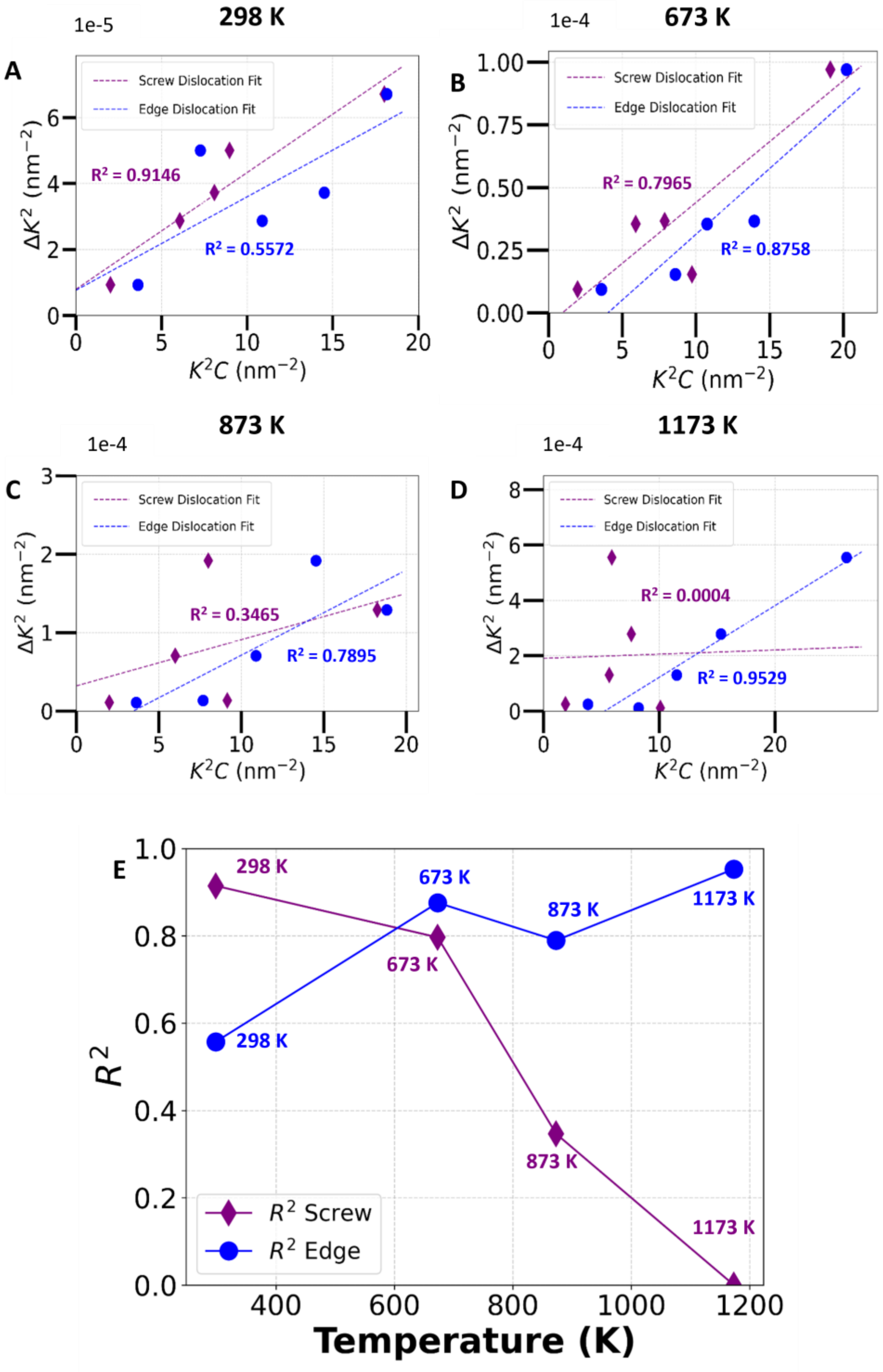

*Figure 4 Modified Williamson–Hall analysis of the NbTaTiV alloy at different temperatures. (A–D) Linear fits of* $(\Delta K)^2$ *as a function of* $K^2C_{hkl}$ *for* $\vec{b} = \frac{a}{2}\langle 111\rangle$ *type screw and edge dislocations at 298 K, 673 K, 873 K, and 1173 K, respectively. (E) Temperature dependence of the goodness of fitting,* $(R^2)$ *for screw and edge dislocation illustrates a gradual transition from screw dominated to edge dominated behavior with increasing temperature.*

Although the temperature-dependent change in goodness of fitting provides compelling qualitative evidence for a screw to edge dislocation transition, quantitative determination of the relative fractions of screw and edge dislocations is not straightforward. Neutron diffraction peak broadening can also arise from other crystallographic defects, crystallite sizes, local lattice distortions, and thermal effects, which complicate direct quantitative interpretation. Consequently, a precise fraction of screw versus edge dislocation densities cannot be defined based on the fitting goodness. Nevertheless, the observed systematic trend remains qualitatively significant and supports the fact that the dominant dislocation character evolves from screw to edge with increasing temperature.

## 4. Conclusions

This study demonstrates a temperature-driven transition from screw- to edge-dislocation-controlled deformation that explains the strong high-temperature strength retention of NbTaTiV. The key findings include:

1. **Low temperature (298–573 K):** Deformation is dominated by straight screw dislocation glide on {110} planes controlled by Peierls barriers and kink-pair nucleation, with strong temperature sensitivity.
2. **Intermediate (573–773 K):** Cross-kink unpinning generates abundant debris; evidence of mixed dislocations at 773 K indicate a transition in dominant character.
3. **High temperature (≥973 K):** Edge dislocation glide dominates; dislocation interactions with V-induced misfit volume reduce edge mobility and weaken temperature dependence of yield strength.

## 5. Acknowledgement

The work was supported by the grant DE-SC0025388 funded by the U.S. Department of Energy, Office of Science. A portion of this study was carried out at the UC Davis Advanced Material Characterization and Testing (AMCaT) Facility. Funding for the Thermo Fisher Quattro S was provided by the National Science Foundation Grant No. MRI-1725618.

Work at the National Center for Electron Microscopy, Molecular Foundry was supported by the Office of Science, Office of Basic Energy Sciences, of the U.S. Department of Energy under Contract No. DE-AC02-05CH11231.

Part of this research utilized resources at the Spallation Neutron Source, a DOE Office of Science User Facility operated by the Oak Ridge National Laboratory. The beam time was allocated to VULCAN Engineering Materials Diffractometer on proposal number IPTS-35022.